\documentclass[prb,twocolumn,footinbib, showpacs,showkeys]{revtex4}
\usepackage{graphicx}
\usepackage{epsfig}
\usepackage{dcolumn}
\usepackage{bm}
\usepackage{amsmath}
\usepackage{amsbsy}
\usepackage{amssymb}
\usepackage{color}

\newcommand{\bra}[1]{\ensuremath{\langle{#1}|\,}}
\newcommand{\ket}[1]{\ensuremath{\,|{#1}\rangle}}

\begin{document}

\title{Electronic transitions in quantum dots and rings induced
by inhomogeneous off-centered light beams}

\author{G.\ F.\ Quinteiro, A.\ O.\ Lucero, and P.\ I.\ Tamborenea}

\affiliation{Departamento de F\'{\i}sica and IFIBA,
Universidad de Buenos Aires, Ciudad Universitaria,
Pabell\'on I, 1428 Ciudad de Buenos Aires, Argentina}

\date{\today}

\begin{abstract}
We theoretically investigate the effect of inhomogeneous light beams
with (twisted light) and without (plane-wave light) orbital angular
momentum on semiconductor-based nanostructures, when the symmetry
axes of the beam and the nanostructure are displaced parallel to
each other.
Exact analytical results are obtained by expanding the off-centered light field
in terms of the appropriate light modes centered around the nanostructure.
We  demonstrate how electronic transitions involving the
transfer of different amounts of orbital angular momentum are
switched on and off as a function of the separation between the axes
of the beam and the system.
In particular, we show that even off-centered plane-wave beams
induce transitions such that the angular momenta of the initial and final
states are different.
\end{abstract}

 \maketitle


FINAL VERSION PUBLISHED IN J. Phys.: Condens. Matter 22 505802 (2010) \\

\section{Introduction}

Over the last few decades, the design and subsequent study of man-made
nanostructures has occupied a crucial place in pure and applied condensed
matter research.
The importance of these systems stems from the fact that, on the one
hand, they are excellent tools to probe the laws of physics at different
length scales, and on the other hand, their technological applications
are vast, spanning electronics, optical devices, and spintronics.
Quantum dots (QDs) are a paradigmatic example that illustrates the
general trend.\cite{rei-man,han-etal}
QDs mimic atoms at a larger length scale (they are called sometimes
artificial atoms) and allow a flexible control of the ``atomic" properties,
like the strength of the Coulomb interaction via the dielectric constant,
the number of electrons, the shape of the confining potential, the spin-orbit
interaction, etc.
Quantum rings (QRs) \cite{war-etal,fuh-etal} are another important example of
nanostructured systems where interesting physical effects can be
explored---Aharonov-Bohm effect, persistent currents, quantum interference,
etc.

Much research is being carried out on semiconductor nanostructures, which
present, among other interesting features, a strong coupling with optical
(or near optical) electromagnetic fields.
Naturally, the electronic states of these structures can be probed and
manipulated using a variety of light pulses.
For instance, the use of femtosecond light pulses is usually claimed to
outperform other methods when an ultrafast control is sought.
However, there has been very little exploration of what can be done on
semiconductor nanostructures by taking advantage of the inhomogeneous
nature of light beams.
In particular, the interaction of these structures with twisted
light (TL)\cite{Andrews-01}---i.e.\ light carrying orbital angular
momentum (OAM)---is only beginning to be studied.
We have recently initiated studies in this direction, both for
QDs~\cite{qui-tam-09b} and QRs.~\cite{qui-ber}

In particular, our previous article in QDs explored the interaction of
disk-shaped QDs with twisted-light, in the case where the beam axis coincides
with the symmetry axis of the QD.
We predicted that the transitions are not vertical and may connect states
having different values of the orbital angular momentum of the electron, and
from an applied standpoint, that the use of light carrying OAM would
facilitate the ultrafast manipulation of the electronic states in
the system.

This article extends our previous work on QDs and QRs to the general
case of the interaction of inhomogeneous fields---with special
emphasis on TL beams---with QDs or QRs when the symmetry axis of the
nanostructure and the beam axis do not coincide. The problem we
study is of interest, for it addresses realistic experimental
situations, like the irradiation of an ensemble of nanostructures,
where each nanostructure will see a displaced light beam. Also, when
a single nanostructure is meant to be addressed, our theory shows
how an imprecise centering of the beam can alter significantly the
optical transitions.
Finally, from a constructive point of view, our
calculations display the wealth of opportunities for optical control
of electronic states which can be achieved by simply changing the
positioning, waist, and orbital angular momentum of the light beam.

The article is organized as follows.
Section~\ref{Sect:Background} briefly covers the already studied problem of
the interaction of TL with electrons in nanostructures (QDs and QRs) when
the beam and system symmetry axes coincide.
Section \ref{Sect:Off-centered case} develops the theory used to solve the
general problem of a TL beam displaced from the symmetry axis of the
nanostructure, and presents our main findings.
Conclusions are given in Sect.~\ref{Sect:Conclusions}.

\section{Background: Centered beam}
\label{Sect:Background}

In this section we treat the problem of the interaction of TL with
QDs and QRs, in the case where both symmetry axes coincide.
The model and results are based on previous studies, and serve as an
introduction to the more general problem of off-centered beams.

Succinctly stated, twisted light is light carrying orbital angular
momentum $\hbar l$, characterized by inhomogeneous fields having a
phase $e^{i l \theta}$ and radial dependence of the Laguerre or
Bessel form.
In the Coulomb gauge, the inhomogeneous vector potential has both transverse
and longitudinal components.
Under typical experimental conditions, the transverse component is the
dominant one, and reads in cylindrical coordinates
\begin{eqnarray} \label{eq:VectorPot_a}
    \mathbf{A}_l(\mathbf{r},t)
&=& \boldsymbol{\epsilon}_\sigma F_{{l}}(\mathbf{{r}})
    e^{i(q_z z-\omega t)}+ \mbox{c.c.} \nonumber \\
&=& \mathbf{A}^{(+)}_l(\mathbf{r},t) +
    \mathbf{A}^{(-)}_l(\mathbf{r},t)\,,
\end{eqnarray}
where $\boldsymbol{\epsilon}_\pm= \hat{x} \pm i \hat{y}$ is the
transverse circular-polarization vector.
We choose to work with Bessel modes for the radial dependence, since these
will allow for analytical results in the next section.
Thus, we take
\begin{equation}
\label{eq:Fdefinition}
    F_{l}(\mathbf{r})
=   J_{{l}}(q_r r )e^{i l \theta}\,.
\end{equation}
%


Quantum dots and rings based on semiconductor materials can be made to
confine both electrons and holes. Electronic states in these structures
can be mathematically described by the product of envelope $R_{}(r)
e^{i m \theta} Z(z) $,  microscopic $u_b(\mathbf{r})$ (here taken at zero
crystal momentum), and spin $\xi$
functions
\begin{equation}
    \Psi_{b}(\mathbf{r})
=   \left[ R_{}(r) e^{i m \theta} Z(z) \right]
    u_b(\mathbf{r}) \, \xi\,.
\end{equation}
In semiconductor systems, strain lifts the degeneracy of the heavy-hole
and light-hole bands.
Therefore, a two-band model in the effective-mass approximation with one
conduction ($b=c$) and one heavy-hole valence ($b=v$) band is sufficient
to treat our problem.
In addition, we have assumed that the envelope function is separable.
Furthermore, since the confinement in the $z$-direction is much stronger
than that in the x-y plane, it is a good approximation to assume that the
electron remains in the lowest-energy $z$-eigenstate.


The coupling between the semiconductor structure and the TL beam is
modelled using the minimal-coupling Hamiltonian, which to lowest order
in the vector potential is given by
\begin{equation}
    h_{l}
=   -\frac{q}{m_e}
    \mathbf{A}_l(\mathbf{r},t) \cdot \mathbf{p}\,,
\end{equation}
where $q=-e$ and $m_e$ are the charge and mass of the electron. This
coupling causes electronic transitions between the conduction and
valence bands.
The physics of the interaction can be derived---e.g.\ using Fermi's
Golden Rule, master equation, Heisenberg equations---from the matrix
elements of $h_{l}$.
Since we are interested in interband transitions, the rotating-wave
approximation is applied and we obtain for the matrix elements of the
light-matter interaction
\begin{eqnarray}
\label{eq:hplus}
    \bra{ c\alpha'} {h}^{(+)}_{l} \ket{v\alpha}
&=& -\frac{2 \pi q}{m_e} e^{-i \omega t}
    \left( \boldsymbol{\epsilon} \cdot \mathbf{p}_{cv} \right)
    \delta_{l,(m'-m)} \times    \nonumber \\
&&  \hspace{-10 mm}
    \delta_{\xi',\xi}
    \int_0^{\infty} dr \,r J_{l}(q_r r )
    R^{*}_{c\alpha'}({r}) R_{v\alpha}({r})\,, \\
\label{eq:hminus}
    \bra{ v\alpha' } {h}^{(-)}_{l} \ket{c\alpha}
&=& -\frac{2 \pi q}{m_e} e^{i \omega t}
    \left( \boldsymbol{\epsilon}^{*} \cdot \mathbf{p}_{vc} \right)
    \delta_{l,(m-m')} \times
    \nonumber \\
&&  \hspace{-10 mm}
    \delta_{\xi',\xi}  \int_0^{\infty} dr \,r J_{l}(q r )
    R^{*}_{v\alpha'}({r}) R_{c\alpha}({r})\,,
\end{eqnarray}
where $\alpha$ is a collective index that gathers all quantum numbers
relevant for QDs (radial and angular) or QRs (angular).
The expressions given above represent absorption ($v \rightarrow c$)
and emission ($c \rightarrow v$) of light, respectively [for a detailed
derivation, see Refs.~(\onlinecite{qui-tam-09b}) and (\onlinecite{qui-ber})].

\section{Off-centered beam}
\label{Sect:Off-centered case}

\begin{figure}[h]
  \centerline{\includegraphics[scale=.9]{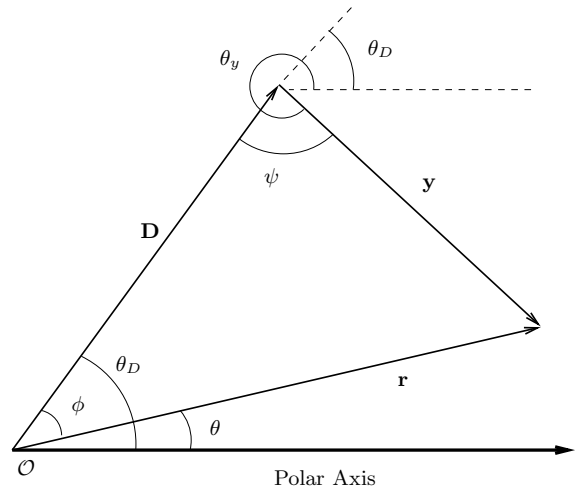}}
  \caption{Relative positions of the nanoparticle and the light beam. The QD/QR is
  placed at the origin $\cal O$, and the TL beam axis passes through $\mathbf D$.
  The $z$-axis is normal to the plane of the drawing.}
  \label{figure:scheme}
\end{figure}

We now consider the case of light beams whose center is displaced with
respect to that of the nanostructure.
We formulate the theory for TL beams with arbitrary OAM, given by $l$;
the case of plane waves is then easily derived by setting $l=0$.
The geometry of the problem is shown in Fig.\ \ref{figure:scheme}.
The nanostructure is centered at the origin of coordinates, with symmetry
axis along $z$, and a beam of TL propagates in the $z$-direction, displaced
from the origin by a distance $D$.
Hereafter, we shall refer to this situation as having an ``off-centered''
TL beam (with respect to the nanostructure).
We first formulate the theory and then analyze the main aspects of its
predictions.
%

\subsection{Theory}

Let us use Eq.~(\ref{eq:VectorPot_a}) to write the vector potential centered
at $\mathbf D$ (hereon, we indicate displaced vector potentials and
Hamiltonians by a tilde)
\begin{equation} \label{Eq:Off-centeredA}
    ‎\tilde{\mathbf{A}}_l(\mathbf{y},t)
=  \boldsymbol{\epsilon}_\sigma F_{{l}}(\mathbf{{y}})
     e^{i(q_z z-\omega t)}+
    \mbox{c.c.}\,,
\end{equation}
where
\begin{eqnarray*}
    F_{l}(\mathbf{y})
&=& J_{{l}}(q_r y ) e^{i l \theta_{y}} \nonumber \\
&=& J_{{l}}(q_r y ) (-1)^l e^{i l \psi} e^{i l \theta_{D}}
\end{eqnarray*}
since $\theta_{y}=\pi + \psi + \theta_{D}$.
The relation~\cite{Korenev}
\begin{equation}
\label{eq:laposta}
    J_{{l}}(q_r y )e^{i l \psi}
=   \sum_{s=-\infty}^{\infty} J_{l+s}(q_r D ) J_{s}(q_r r ) e^{is{\phi}}
\end{equation}
allows us to express $F_{l}(\mathbf{y})$ in terms of centered coordinates
as
\begin{eqnarray*}
    F_{l}(\mathbf{y})
&=& (-1)^{l} \!\! \sum_{s=-\infty}^{\infty} \!\! J_{l+s}(q_r D )
    J_{s}(q_r r ) e^{i s{(\theta_{D}-\theta)}} e^{i l \theta_{D}} \\
&=& \sum_{s=-\infty}^{\infty} (-1)^{l-s} F_{l-s}(\mathbf{D})
    F_{s}(\mathbf{r})\,.
\end{eqnarray*}
\begin{figure}[h]
  \centerline{\includegraphics[scale=1]{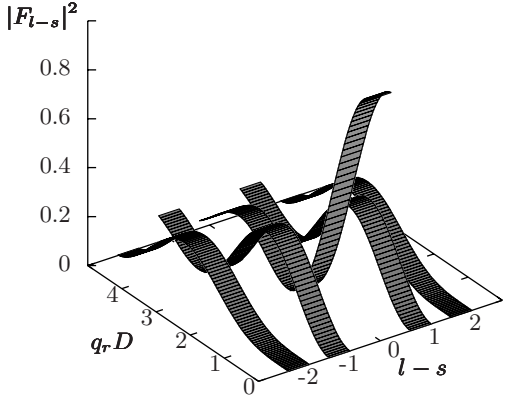}}
  \caption{Weights corresponding to the decomposition of the off-centered
  TL beam in terms of TL beams centered at the origin.}
  \label{fig:cintas}
\end{figure}
From Eq.~(\ref{Eq:Off-centeredA}), the final form of the vector
potential centered at position $\mathbf D$ is
\begin{eqnarray}
    \tilde{\mathbf{A}}_l(\mathbf{y},t)
&=& \sum_{s=-\infty}^{\infty}(-1)^{l-s}F_{l-s}(\mathbf{D})
    \, \mathbf{A}^{(+)}_s(\mathbf{r},t) + \nonumber \\
&&  \hspace{-0 mm}
    \sum_{s=-\infty}^{\infty}(-1)^{l-s}F^{*}_{l-s}(\mathbf{D})
    \, \mathbf{A}^{(-)}_s(\mathbf{r},t)
    \label{eq:superposition}
\end{eqnarray}
is given in terms of a superposition of TL vector potentials centered at
the origin, each having different OAM, and weighed by
$F_{l-s}(\mathbf{D})$.
This expansion is easily understood by analogy with the transformation
of angular momentum in classical mechanics, when the proper axis---from
which the motion is seen as a pure rotation---is changed to an arbitrary axis.
Although our transformation makes the representation of
$\tilde{\mathbf{A}}_l(\mathbf{y},t)$ more complex, it is justified by the need
to refer both the TL beam and the electronic states to the same axis.
Figure \ref{fig:cintas} shows the weights squared
$|F_{l-s}(\mathbf{D})|^2= J_{l-s}(q_r D)^2$ of the decomposition of
Eq.~(\ref{eq:superposition}).

It is now a simple algebraic matter to write down the interaction
Hamiltonian $\tilde{h}_{l}= (-q/m_e)
\tilde{\mathbf{A}}_l(\mathbf{y},t) \cdot \mathbf{p}$, which becomes
\begin{eqnarray*}
    \tilde{{h}}_{l}
&=& \sum_{s=-\infty}^{\infty}(-1)^{l-s}F_{l-s}(\mathbf{D})
    \left[-\frac{q}{m_e} \mathbf{A}^{(+)}_s(\mathbf{r},t)
    \cdot \mathbf{p}\right]
    + \nonumber \\
&&  \hspace{-0 mm}
    \sum_{s=-\infty}^{\infty}(-1)^{l-s}F^{*}_{l-s}(\mathbf{D})
    \left[-\frac{q}{m_e} \mathbf{A}^{(-)}_s(\mathbf{r},t)
    \cdot \mathbf{p}\right] \,,
\end{eqnarray*}
or, using the the Hamiltonian at the origin $ {h}_{l}$
\begin{eqnarray}
    \label{eq:H}
    \tilde{{h}}_{l}
&=& \sum_{s=-\infty}^{\infty}(-1)^{l-s}F_{l-s}(\mathbf{D})
    {h}^{(+)}_{s} + \nonumber \\
&&  \hspace{-0 mm}
    \sum_{s=-\infty}^{\infty}(-1)^{l-s}F^{*}_{l-s}(\mathbf{D})
    {h}^{(-)}_{s} + \nonumber \\
&=& \hspace{-0 mm}
    \tilde{h}^{(+)}_{l} + \tilde{h}^{(-)}_{l} \,.
\end{eqnarray}
The matrix elements of the Hamiltonian, Eq.~(\ref{eq:H}), are
determined using the matrix elements for the centered case,
Eqs.~(\ref{eq:hplus}) and (\ref{eq:hminus}), and give
\begin{eqnarray}
        \bra{ c\alpha'} \tilde{h}^{(+)}_{l} \ket{v\alpha}
&=&
        \sum_{s=-\infty}^{\infty}(-1)^{l-s}F_{l-s}(\mathbf{D})
        \bra{c\alpha'} {h}^{(+)}_{s} \ket{v\alpha}  \nonumber \\
        \bra{  v\alpha'} \tilde{h}^{(-)}_{l} \ket{c\alpha}
&=&
        \sum_{s=-\infty}^{\infty}(-1)^{l-s}F^{*}_{l-s}(\mathbf{D})
        \bra{v\alpha'} {h}^{(-)}_{s} \ket{c\alpha} \nonumber \,.
\end{eqnarray}
The sums are solved thanks to the delta function
$\delta_{s,\pm(m'-m)}$ that appear in the matrix elements of $h_s^{(\pm)}$.
Then,
\begin{eqnarray}
    \bra{ c\alpha'} \tilde{h}^{(+)}_{l} \ket{v\alpha}
&=& (-1)^{l-(m'-m)}F_{l-(m'-m)}(\mathbf{D})
    \nonumber \\
&&  \hspace{-0 mm} \bra{v\alpha'} {h}^{(+)}_{m'-m}
   \ket{c\alpha} \\
    \bra{ v\alpha'} \tilde{h}^{(-)}_{l} \ket{c\alpha}
&=& (-1)^{l-(m-m')}
    F^{*}_{l-(m-m')}(\mathbf{D}) \nonumber \\
&&  \hspace{-0 mm}
    \bra{v\alpha'} {h}^{(-)}_{m-m'} \ket{c\alpha}\,.
\end{eqnarray}
The final expressions read
\begin{widetext}
\begin{eqnarray}\label{eq:ME_Hplus}
    \bra{  c\alpha'} \tilde{h}^{(+)}_{l} \ket{v\alpha}
&=& - \frac{2 \pi q}{m_e} e^{-i \omega t}
    \left( \boldsymbol{\epsilon}_\sigma \cdot  \mathbf{p}_{cv} \right)
    F_{l-(m'-m)}(\mathbf{{D}})
   (-1)^{l-(m'-m)}
    \int_0^{\infty} dr \,r J_{m'-m}(q r )
    R^{*}_{c\alpha'}({r}) R_{v\alpha}({r})\,,  \\
\label{eq:ME_Hminus}
\bra{  v\alpha'} \tilde{h}^{(-)}_{l} \ket{c\alpha}
&=& - \frac{2 \pi q}{m_e} e^{i \omega t}
    \left(\boldsymbol{\epsilon}^{*}_\sigma \cdot \mathbf{p}_{vc} \right)
    F^{*}_{l-(m-m')}(\mathbf{{D}})
    (-1)^{l-(m-m')}
    \int_0^{\infty} dr \, r J_{m-m'}(q r ) R^{*}_{v\alpha'}({r})
    R_{c\alpha}({r})\,,
\end{eqnarray}
\end{widetext}
where, to ease the notation, we eliminated the delta functions for
the spin indices, and assumed that the bands are the correct ones which
yield non-vanishing matrix elements.

\subsection{Results and analysis}

We would like to draw first a general picture of what happens to
the electronic transitions when the TL beam is displaced.
To this end, let us consider the main ingredient in the calculation of the
transition probability using Fermi's Golden Rule, namely
\begin{equation}
\label{eq:trans}
    \left|\bra{v\alpha'} \tilde{h}^{(+)}_{l} \ket{c\alpha}\right|^2
= \kappa \left|F_{n}(\mathbf{D})\right|^2
= \kappa J_n(q_r D)^2 \,,
\end{equation}
where $n=l-(m'-m)$.
Note that $\kappa$ contains relevant information, that will be unfolded
later in this section, when precise numerical calculations are presented.

\begin{figure}[h]
  \includegraphics[scale=.9]{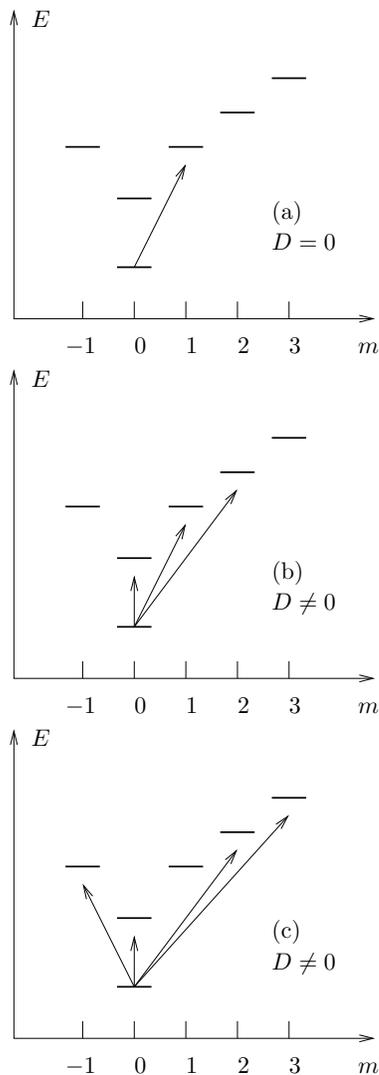}
  \caption{Pictorial representation of the possible transitions in a
  QR, induced by TL having OAM $l=1$.
  $(a)$ A centered beam produces a non-vertical transition between
  a valence-band and a conduction-band state differing in one unit
  of angular momentum $m$.
  $(b)$ An off-centered beam produces several non-vertical transitions.
  $c)$ For a particular choice of $q_r D$ the transition from the upper
  valence-band state to the $m=1$ conduction-band state becomes negligible. }
  \label{fig:trans}
\end{figure}

For concreteness, let us assume that the TL beam possesses $l=1$; it
is worth stressing that $l$ is the OAM of the beam as seen from its
proper axis.
When the beam is centered, $ J_n(q_r D=0)^2 \neq 0$ only for $n=0$.
This implies that the only allowed transition is that connecting
states that differ by one unit of angular momentum: $m'-m=1$, see
Fig.\ \ref{fig:trans}(a).
As the beam is displaced off the nanostructure symmetry axis, other
transitions between states having $m'-m = \pm 2, \pm 3, \ldots$
become allowed, see Fig.\ \ref{fig:trans}(b).
To understand this apparent violation of conservation of OAM, we recall
that, according to Eq.~(\ref{eq:superposition}), the off-centered beam
is a superposition of centered beams with varying values of OAM.
Due to the $n$-dependence of the zeros of the Bessel function,
there are specific values of $q_r D$ such that, for instance, the
transition matrix element between states differing by one unit of
angular momentum is zero, while the probability for other
transitions remains finite, as illustrated in Fig.\ \ref{fig:trans}(c).

%


In order to perform a numerical analysis, we shall now consider the
whole expressions in Eqs.~(\ref{eq:ME_Hplus}) and (\ref{eq:ME_Hminus}).
Thus, we need to consider specific radial functions.
For the case of a QR
\begin{eqnarray}
        R(r)
&=& \sqrt{2 \over r_0 d} \sin{\left[ \frac{\pi}{d}(r-r_0+d/2)
        \right]}
    \,,
\end{eqnarray}
with $r_0$ and  $d$ the radius and width of the ring; $R(r)=0$
outside the ring. We assume $r_0 \gg d$, which allows us to restrict
the analysis to one subband. For a QD, the radial wave function
reads
\begin{eqnarray}
    R_{sm}(r)
&=&
    \frac{(-1)^s}{\sqrt{2\,\pi}\,\ell}
    \,\sqrt{\frac{s!}{(s+|m|)!}}e^{-{r^2/(4\,\ell^2)}}
   \times \nonumber \\
&&
     \left( \frac{r}{\sqrt{2}\,\ell} \right)^{|m|}
     L_s^{|m|} \left(r^2/(2\,\ell\,^2)\right)
    \,,
\end{eqnarray}
where $\ell$ is a characteristic length for the confinement of
electrons, $s$ and $m$ are the radial and angular momentum quantum
numbers, and $L_s^{|m|} \left(x\right)$ is a Laguerre polynomial.

For the QR, the integral reads---the quantum number indices in
$R(r)$ are unnecessary---
\begin{eqnarray}
    \int_0^{\infty} dr \,r J_{m'-m}(q r )
    |R({r})|^2
&=&   J_{m'-m}(q_r r_0 )
    \,.
\end{eqnarray}
For the case of a QD, a redefinition of the radial coordinate to $x
= r^2/(2\ell^2)$, allows to simplify the expression.\cite{qui-tam-09b}
In the following, we exemplify by using the case of a QR. The
matrix element for the absorption of light becomes
\begin{eqnarray}\label{eq:ME_Hplus_QRing}
        \bra{  c\alpha'} \tilde{h}^{(+)}_{l} \ket{v\alpha}
&=& -  (-1)^{n} \frac{2 \pi q}{m_e}
     \left( \boldsymbol{\epsilon}_\sigma
        \cdot   \mathbf{p}_{cv} \right) \times \nonumber \\
&&  F_{n}(\mathbf{{D}})
    J_{m'-m}(q_r r_0 ) e^{-i \omega t}
   \,.
\end{eqnarray}
We are now in a position to precisely determine the transition
probability. We see at once that the probability for the transition
$\{v\alpha\} \rightarrow \{c\alpha'\}$ is determined by the
displacement $\mathbf D$, the difference $m'-m$ between the angular
momentum quantum numbers, and the relative sizes of QR ($r_0$) to
beam waist ($\simeq q_r^{-1}$).

For a general displacement $D\neq 0$, several transition matrix
elements have finite values, as explained before. The relative
strength of each transition is determined by $ [J_{l-(m'-m)}(q_r D)
J_{m'-m}(q_r r_0)]^2$.

\begin{figure}[h]
  \includegraphics[scale=1.]{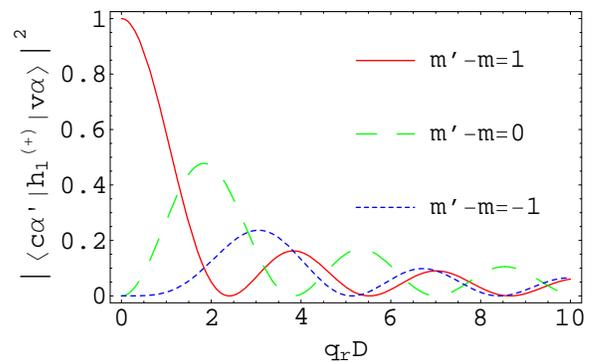}
  \caption{ (color online)
            Normalized transition matrix element for a TL beam with
            $l=1$ and for different values of the initial and final
            angular momentum quantum number. }
  \label{fig:curves}
\end{figure}

Let us exemplify for a TL beam with $l=1$; then, we have
$n=1-(m'-m)$.
Figure~\ref{fig:curves} shows the quantity
$[J_{1-(m'-m)}(q_r {D}) J_{m'-m}(q_r r_0)]^2$
for different values of $D$ and $m'-m$.
We observe what was said previously from a qualitative point of view.
At zero displacement, the only contribution comes from transitions
differing by one unit in their angular momentum.
When the beam is displaced, other transitions become allowed.
An extinction  occurs when the value $q_r {D}$ is a
zero of the Bessel function.

\begin{figure}[h]
  \includegraphics[scale=1.]{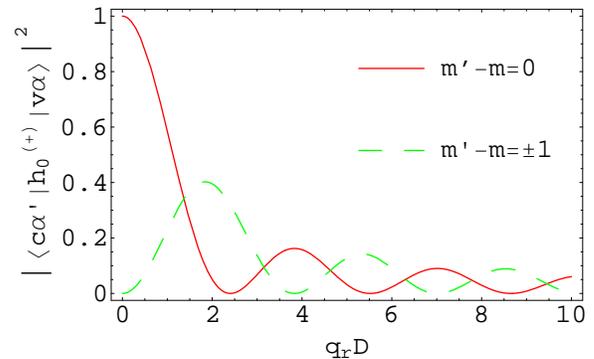}
  \caption{(color online)
            Normalized transition matrix element for plane-wave light
            ($l=0$) and for different values of the initial and final
            angular momentum quantum number. }
  \label{fig:curves2}
\end{figure}

Another interesting situation arises when the beam carries no OAM,
i.e.\ $l=0$ in Eq.\ (\ref{eq:ME_Hplus_QRing})---we refer to this as
the plane wave situation. In Fig. \ref{fig:curves2} where we analyze
$[J_{m'-m}(q_r {D}) J_{m'-m}(q_r r_0)]^2$. As expected, when the
beam is centered on the nanoparticle, the only allowed transitions
are vertical, i.e. $\{v\alpha\} \rightarrow \{c\alpha\}$; however,
when the beam axis is displaced, non-vertical transitions become
allowed. An important difference arises with respect to the $l\neq
0$ case; since the function $[J_{m'-m}(q_r {D}) J_{m'-m}(q_r
r_0)]^2$ is symmetric on $m'-m$, the transitions having $m'-m=\pm 1,
\pm 2, \ldots$ have equal probability, and in most situations there
will be no net transfer of orbital angular momentum. However, this
symmetry can be broken by the application of a magnetic field, which
detunes one of the possible final states. Also, this symmetry is
broken (even without an applied magnetic field) if the initial
population of valence-band states is asymmetric with respect to the
quantum number $m$. The peculiar effects that we predict are the
result of the finite waist of the beam, and the fact that, to treat
it adequately, we have gone beyond the dipole-moment approximation.

The relative size of QR to beam waist impacts on the response of
QRs differing in their radii $r_0$, but lying at the same distance
$D$ from the beam axis---this situation may arise, for example, due
to the statistical nature of the fabrication process of an ensemble
of nanoparticles.
As a consequence of the factor $J_{m'-m}(q_r r_0)$
in Eq.~(\ref{eq:ME_Hplus_QRing}), the various electronic transitions
are enhanced and inhibited to some degree, and this results in a
different evolution of the electronic states in each QR.

\section{Conclusions}
\label{Sect:Conclusions}

We have theoretically investigated the effect that an inhomogeneous
light beam, such as plane waves ($l=0$) with a finite waist or
twisted light beams ($l \neq 0$), has on semiconductor-based nanostructures,
when the symmetry axes of the beam and the nanoparticle do not coincide.

The problem was analytically solved for the general case of twisted light
with $l=0,1,2,\ldots$ by writing the off-centered beam as a superposition
of twisted light beams centered at the position of the nanoparticle.
This decomposition allowed us to study the possible electronic
transitions in terms of the problem of a centered beam illuminating
a nanoparticle, which we have already investigated in a previous
work.

We showed that different transitions between states in the
nanoparticle are switched on and off as a function of the distance
from the beam to the nanoparticle axes. In addition, the strength of
the transition is determined by the relative size of the
nanostructure to the beam waist. Our results also predict that a
plane-wave beam with a finite waist will induce both vertical and
non-vertical transitions depending on the displacement of the axes.

Our study indicates that, under several experimental conditions, care must
be taken when interpreting the results of light-nanoparticle interaction.
For example, if an ensemble of nanostructures is illuminated in normal
incidence using a beam whose waist is roughly of the same size or smaller
than the ensemble, each nanoparticle will respond in a different way,
irrespective of the beam being a plane wave or twisted light.
In a situation where the experiment is carried out on a single nanostructure,
we saw that the precise positioning of the TL axis is crucial to know
what transitions are being excited.

\section{Acknowledgement}

This research was financially supported by grant PICT-02134/2006
from Agencia Nacional de Promoci\'on Cient\'ifica y Tecnol\'ogica and by
grant UBACYT X495.


\end{document}